\documentclass[conference]{IEEEtran}
\IEEEoverridecommandlockouts
\usepackage{cite}
\usepackage{amsmath,amssymb,amsfonts}
\usepackage{algorithmic}
\usepackage{graphicx}
\usepackage{textcomp}
\usepackage{xcolor}
\def\BibTeX{{\rm B\kern-.05em{\sc i\kern-.025em b}\kern-.08em
    T\kern-.1667em\lower.7ex\hbox{E}\kern-.125emX}}
\begin{document}

This work has been submitted to Wiley for possible publication. Copyright may be transferred without notice, after which this version may no longer be accessible.

\newpage

\title{Machine learning for accelerating the discovery of high performance low-cost solar cells: \\ a systematic review\\
}

\author{\IEEEauthorblockN{Satyam Bhatti\IEEEauthorrefmark{1}, Habib Ullah Manzoor\IEEEauthorrefmark{1}, Bruno Michel\IEEEauthorrefmark{2}, Ruy Sebastian Bonilla\IEEEauthorrefmark{3}, Richard Abrams\IEEEauthorrefmark{4}, \\ Ahmed Zoha\IEEEauthorrefmark{1}, 
Sajjad Hussain\IEEEauthorrefmark{1} and Rami Ghannam\IEEEauthorrefmark{1}}
\IEEEauthorblockA{\textit{\IEEEauthorrefmark{1}James Watt School of Engineering, University of Glasgow, United Kingdom} \\
\textit{\IEEEauthorrefmark{2}IBM Research GmbH, Zurich, Switzerland}\\
\textit{\IEEEauthorrefmark{3}Department of Materials, University of Oxford, United Kingdom}\\
\textit{\IEEEauthorrefmark{4}Onshore Renewables, AqualisBraemar LOC Group, United Kingdom}\\
\textit{Email: s.bhatti.2\@research.gla.ac.uk}, \textit{rami.ghannam\@glasgow.ac.uk}}
}

\maketitle

\begin{abstract}
Solar photovoltaic (PV) technology has merged as an efficient and versatile method for converting the Sun’s vast energy into electricity. Innovation in developing new materials and solar cell architectures is required to ensure lightweight, portable, and flexible miniaturized electronic devices operate for long periods with reduced battery demand. Recent advances in biomedical implantable and wearable devices have coincided with a growing interest in efficient energy-harvesting solutions. Such devices primarily rely on rechargeable batteries to satisfy their energy needs. Moreover, Artificial Intelligence (AI) and Machine Learning (ML) techniques are touted as game changers in energy harvesting, especially in solar energy materials. In this article, we systematically review a range of ML techniques for optimizing the performance of low-cost solar cells for miniaturized electronic devices. Our systematic review reveals that these ML techniques can expedite the discovery of new solar cell materials and architectures. In particular, this review covers a broad range of ML techniques targeted at producing low-cost solar cells. Moreover, we present a new method of classifying the literature according to data synthesis, ML algorithms, optimization, and fabrication process. In addition, our review reveals that the Gaussian Process Regression (GPR) ML technique with Bayesian Optimization (BO) enables the design of the most promising low-solar cell architecture. Therefore, our review is a critical evaluation of existing ML techniques and is presented to guide researchers in discovering the next generation of low-cost solar cells using ML techniques.
\end{abstract}


\section{Introduction}
Current miniature portable and implantable devices rely on batteries that need replacement and are hazardous to patients. \cite{mukherjee2022state, zhao2019emerging,liu2021piezoelectric} Surgical removal is required when replacing  batteries in implantable devices, which may be inconvenient for patients. \cite{hannan2014energy, zhao2020self} Moreover, implantable biomedical devices are often powered using wires, which may cause discomfort, skin infections, and other hazards to patients. \cite{8960457} The key issues with implanting batteries include metal poisoning for patients due to battery degradation, thus leading to malfunction in generating signals and the damage of electronic circuits. \cite{de1999pacemaker}

Due to their high energy density, scavenging solar energy using photovoltaic (PV) cells has emerged as a potential and feasible solution to power miniature portable and implantable devices. \cite{Zhao2020_PVImplant, stach2021autonomous, wahba2020prediction, Zhao2021_PVSimulation} In general, the architecture of these solar cells can be designed as regular, inverted, mesoporous or planar structures. Furthermore, solar cells combine various materials to enable efficient photon absorption, electron transport, and electron extraction to an external circuit. This means there are vast opportunities for discovering solar cell materials and architectures. In fact, solar cell fabrication techniques involve optimizing different coating materials, thermal annealing conditions, encapsulation methods, etc., which often takes place in the research laboratory. \cite{kumar2022opportunities} However, despite their benefits, these harvesters still suffer from poor efficiency, weak stability, rigidity, and a relatively high cost. \cite{bhuvaneswari2009solar} Promising PV technologies that aim to overcome issues with rigidity and high cost include Perovskite Solar Cells (PSC), Organic Solar Cells (OSC), and Dye-Sensitized Solar Cells (DSSCs). \cite{jacoby2016future} Despite rapid progress in the PSC and OSC field, the stability and efficiency of these low-cost, thin-film solar cells are still poor due to the effects of moisture and temperature. \cite{beard2014promise}. Consequently, machine learning (ML) and artificial intelligence (AI) can be used to improve the performance and accelerate the discovery of these low-cost solar cells \cite{Ghannam_2019}.
 
Innovation in developing new low-cost solar cells is needed, which can be achieved with the help of experimentally validated finite element modelling using software tools such as Sentaurus TCAD. However, this is a time-consuming effort, and leveraging the power of AI can be a game changer in discovering new materials and fabrication techniques to help expedite the process of selection, design, and optimization. \cite{odabacsi2019performance} In fact, the literature suggests that low-cost thin-film solar cell performance can be optimized using a variety of efficient computational and statistical methods. \cite{yilmaz2021critical} From the systems perspective, ML algorithms can also help develop reconfigurable PV cells based on switchable CMOS addressable switches. \cite{zhao2020self}

In the literature, ML relates to the development and ability of the model to learn to adapt, forecast, and predict the independent variables.\cite{el2015machine} ML algorithms consist of 3 types, namely, Supervised learning, Unsupervised learning and Reinforcement learning. \cite{wang2016machine} The supervised ML takes the input data from the user to learn from past experiences and, accordingly, trains the model. \cite{shavlik1990readings} However, the unsupervised ML train model depends upon the real-time data generated and outputs depending on the information given by the user. In contrast, reinforcement learning is the subset of ML that enables an AI-driven system (also known as an agent) to learn by performing tasks and receiving feedback from its trials and errors. \cite{provost1998applied} Herein, we discuss the various ML techniques in-depth that are applied to find an optimized structure for solar cells.

Examples of ML techniques reported in the literature include linear regression, logistic regression, k-nearest neighbours (KNN), random forest (RF), etc., \cite{jordan2015machine, shalev2014understanding} however; every problem requires a unique ML algorithm. \cite{wolpert1997no} Every algorithm has unique abilities and data requirements. For instance, due to nonlinear relations in solar cells, linear regression would not be very helpful. For logistic regression, we have to assume that factors are independent of each other, which might not be the case in solar cells. Similarly, the purpose of KNN is to locate the nearest neighbours with the best possible value. However, it is more suitable for continuous variables. So, the use of ML in optimizing solar cells depends upon the type of experiment, optimizing variables, and data type.

Since the fabrication of OSCs is cheap, most experimental work is carried out via trial and error, which does not guarantee the best performance. \cite{oviedo2020bridging} Instead, researchers are now turning their attention to data-driven techniques for material design and discovery. \cite{mahmood2022machine} ML is one of the vital data-driven techniques that is rising to prominence in discovering new solar cells, forecasting electrical characteristics, and performance prediction without any experimentation. \cite{zhang2020machine, butler2018machine} ML uses algorithms to visualize and analyze data that has several advantages over traditional programming techniques. \cite{parikh2022machine} This paper reviews the different ML algorithms used to find an optimized structure of a low-cost solar cell. The output power can be optimized for different light conditions and shading depending on the positioning of the solar cells. \cite{workman2020machine} In our paper, we discuss the integration of ML methods for designing low-cost solar cells and, consecutively, explore the literature on using different ML techniques for the advanced discovery of solar cells.

\subsection{Contributions to the literature}

In our systematic review, we analyzed the role of ML in the field of solar cell design and material discovery. We conducted a systematic review of the applications of ML in the optimization, fabrication, and discovery of new photovoltaic materials. Our article is the first effort to provide a systematic review in this domain. The following are the major contributions of this article:

\begin{figure*}[ht]
 \centering
 \includegraphics[width=18cm]{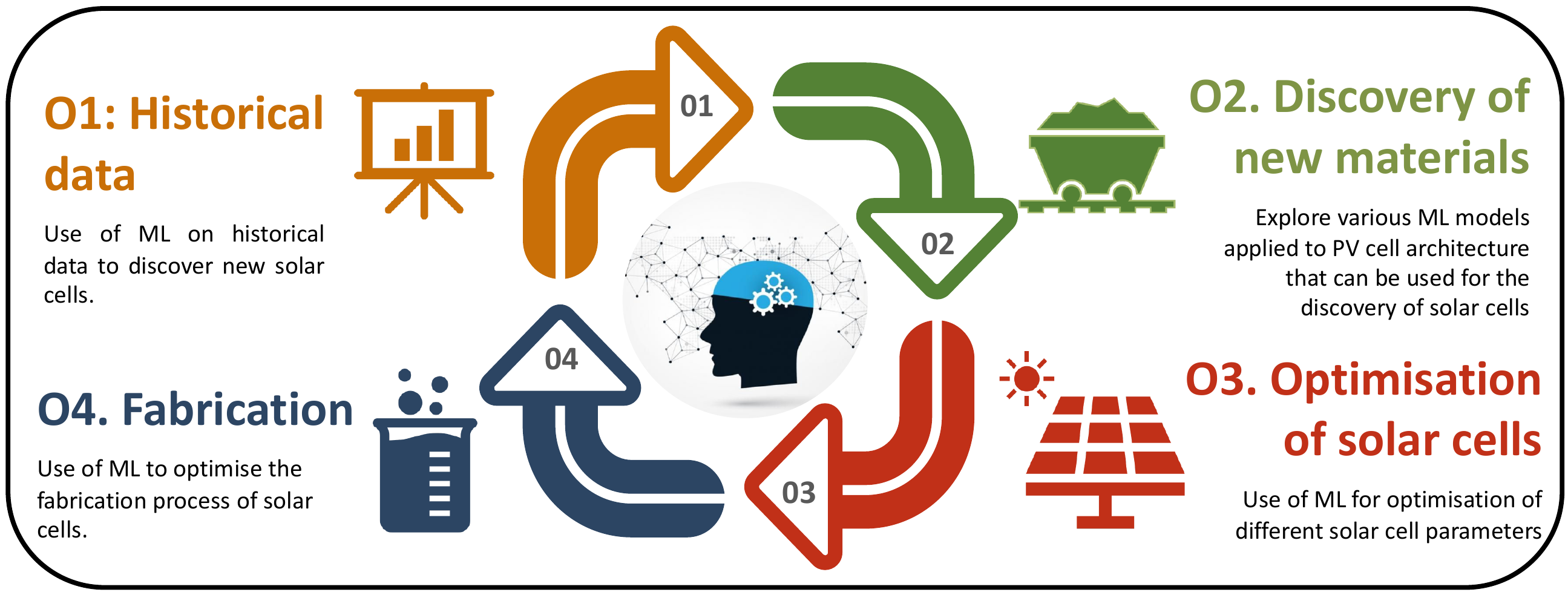}
 \caption{\emph{The objectives of our research are fourfold. The first objective, O1, involves identifying all the literature on low-cost solar cell designs using ML techniques. Our second objective, O2, involves reviewing the literature on materials discovery, whereas O3 identifies specific ML techniques used for optimizing solar cell architectures. Lastly, O4 involves classifying the range of ML algorithms for designing low-cost PV cells from circuits and systems perspective.}}
 \label{fig:Ref1}
\end{figure*}

\begin{enumerate}
    \item We conduct a review of 58 papers from a total of 18,380 research articles involving solar cell discovery, optimization, and fabrication using ML techniques.
    \item We shortlist all ML models that can help in the discovery of new materials.
    \item We review the literature on low-cost high-performance solar cells using ML techniques.
    \item Various ML techniques facilitating the discovery of solar cells were considered in the study.
    \item We investigate the techniques used for the optimization of solar cells with the help of ML.
    \item We highlight the challenges associated with using ML techniques for solar cell design.
\end{enumerate}

\subsection{State of the Art}
During the past 5 years, there has been a surge in the use of ML and AI techniques for designing new solar cells. \cite{parikh2020solar, bash2022accelerated} In this subsection, we review previously published systematic review papers on this field using ML techniques, and we discuss their limitations as well as the contributions that this review provides to the literature.

Qiuling \textit{et al.} \cite{tao2021machine} reviewed the ML techniques for only perovskite materials design and discovery. However, their review lacks a comprehensive comparison of ML techniques for other low-cost solar cells, such as organic, inorganic, hybrid, and DSSCs. Additionally, Hannes \textit{\textit{et al.}} \cite{michaels2021challenges} discussed the challenges of ambient hybrid solar cells for IoT devices, while the paper presented by Hannes \textit{et al.} \cite{wagner2021machine} reveals the study on solar cell cracks using statistical parameters of electroluminescent images using ML. However, both studies presented limited ML algorithms to explore solar cell electrical characteristics. 

Furthermore, Yongjie \textit{et al.} \cite{cui2020recent} reviewed recent advances in computational chemistry for OSC discovery and mentioned the DFT, time-dependent DFT, all atomic molecular dynamics and coarse-grained molecular dynamics. Although their review covered OSCs, it lacked the ML techniques to expedite the process. Next, Florian \textit{et al.} \cite{hase2020designing} reviewed the literature on designing light-harvesting devices using ML, but the review was limited to only OSCs. Likewise, a review paper presented by Sheng \textit{et al.} \cite{jiang2021machine} covered only  ML optimization of PCSs. The studies presented by Anton \textit{et al.}\cite{oliynyk2019virtual}, Min-Hsuan \textit{et al.} \cite{lee2020performance}, and Cagla \textit{et al.} \cite{odabacsi2019performance} explored ML approaches to discover solar cell performance analysis. However, a major drawback in these studies was that limited ML approaches were discussed and did not involve the scope for optimization as well as the fabrication of solar cells in the real environment. 

Therefore, based on the above, state-of-the-art review articles on ML for solar cell discovery focused mainly on a single ML technique with a set of input data. In this work, we instead aim to systematically review the range of ML techniques for developing solar cells. These ML techniques include the procedure to pre-process the input data, various ML algorithms, optimization, and fabrication of the solar cell in a real environment. In this context, our systematic review goes beyond existing literature as it showcases how various ML techniques can accelerate the discovery of high-performance, low-cost solar cells.

\begin{figure*}[ht]
 \centering
 \includegraphics[width=18cm]{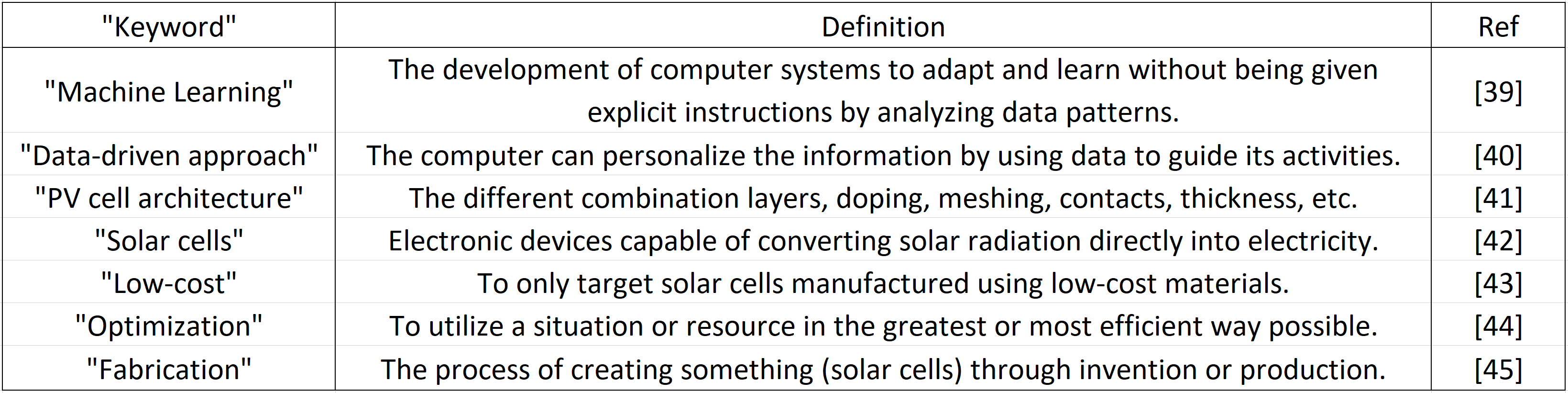}
 \caption{\emph{Keywords and their definitions used for our search from January'2018 to August'2022.}}
 \label{fig:Keywords}
\end{figure*}

\subsection{Organisation of the Article}
The rest of the paper is organized as follows. The adopted methodology in reviewing the literature is discussed in section 2, and the overall results of our systematic review in response to our research questions are presented in section 3. In Section 4 we discuss areas of further study, future outlook, recommendations and open research issues. Finally, summarising remarks are included in the conclusions section.

\section{Review Methodology}

In this section, we discuss our research objectives and our methodology in collecting and synthesizing the literature on ML algorithms for designing and fabricating low-cost, high-performance solar cells.

\subsection{Research objectives}

The four key objectives of our systematic review article are:

\begin{enumerate}
    \item [O1:] To review the range of ML techniques for designing low-cost solar cells using historical data.
    \item [O2:] To identify the ML techniques used specifically for the discovery of new PV materials.
    \item [O3:] From a device perspective, identify the specific ML and optimization techniques used for designing efficient solar cell architectures. 
    \item [O4:] To identify ML algorithms specifically used for the fabrication of low-cost PV cells from the circuits and systems perspective. 
\end{enumerate}

Figure 1 maps our four research objectives and the process involved in shortlisting the research articles. Initially, we focused on extracting and pre-processing the historical data, followed by the discovery of new materials and optimization of solar cells. Lastly, we reviewed the research articles that discussed the integration of ML for fabricating solar cells. Accordingly, in our systematic review, we defined these research objectives to target a set of questions that are the need for the study. Additionally, we shortlisted a set of research articles using the search engines available on Google for extracting the recent research articles published in this domain. This search was subsequently validated using the IBM Watson Studio tool. 

\subsection{Research questions}
Our systematic review aims to answer the following four research questions:

\begin{enumerate}
  \item [RQ1:] What are the data-driven approaches for designing low-cost high-performance solar cells?
  \item [RQ2:] How can ML algorithms facilitate the discovery of new low-cost solar cell materials?
  \item [RQ3:] What are the optimization techniques used for designing an efficient low-cost solar cell architecture?
  \item [RQ4:] What ML algorithms are used for fabricating low-cost solar cells from a circuits and systems perspective? 
\end{enumerate}

\subsection{Review protocol}
For structuring our systematic review, we instigated a review protocol, and the following are the perquisites of the adopted analogy. In this section, we discuss the search strategy, inclusion criteria, exclusion criteria, and screening mechanisms for selecting relevant research papers.

\begin{figure*}[ht]
 \centering
 \includegraphics[width=12cm]{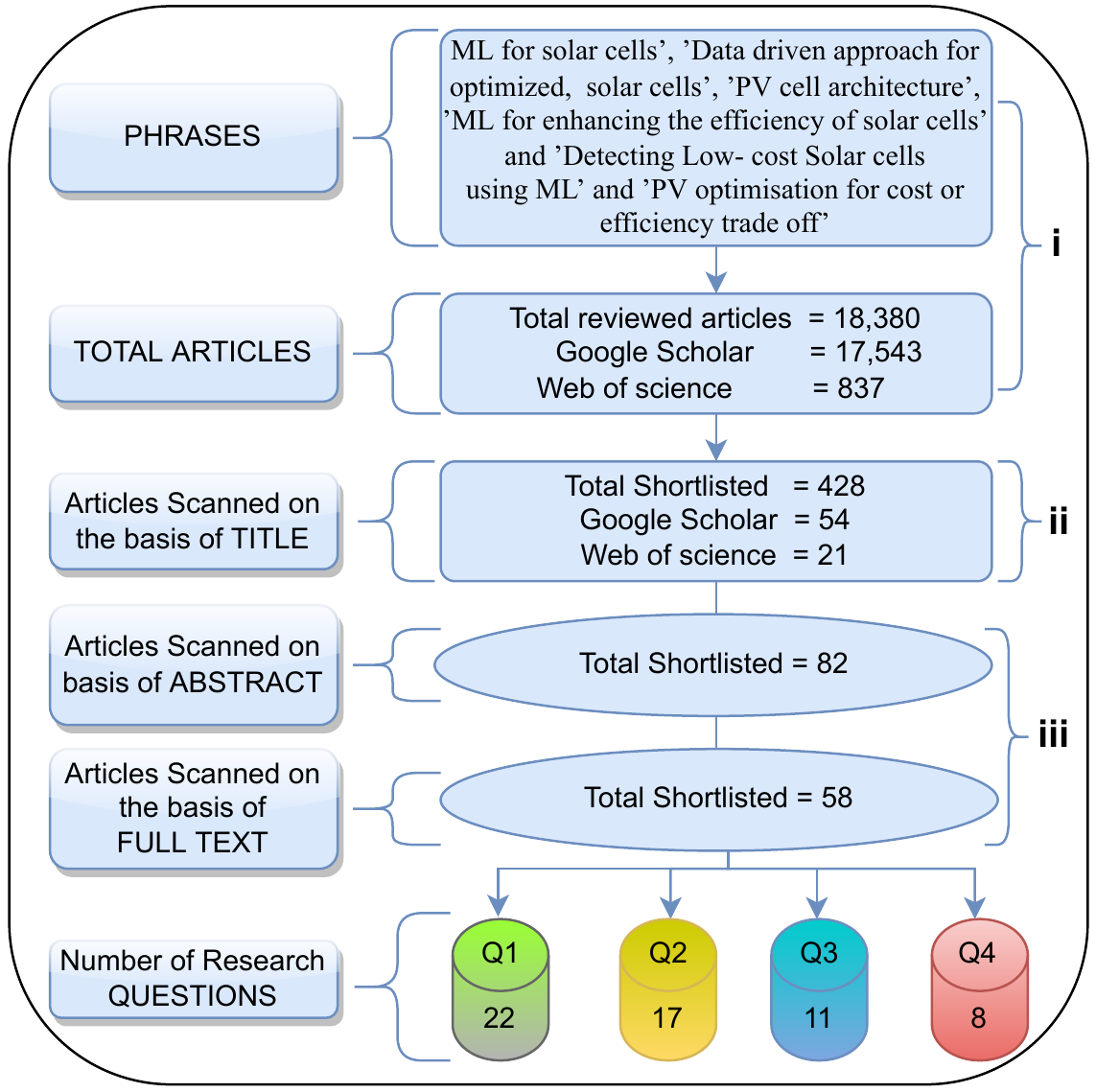}
 \caption{\emph{The PRISMA model represents the process of shortlisting the research articles, including the screening phase based on our assigned research questions from January 2018 to August 2022. The screening of the research articles was done on search engines such as Google Scholar and Web of Science. The respective combination of keywords and phrases were added to the advanced search and subsequently, the articles were shortlisted from manual screening. Further, the total research articles were manually screened based on reading the title, abstract, and full text of the research papers. Therefore, the four questions, Q1, Q2, Q3, and Q4, resulted in a total of 22, 17, 11, and 8 research articles, respectively.}}
 \label{fig:Ref2}
\end{figure*}

\subsubsection{Search strategy}
Our review considered the latest research articles from major publishing houses that include IET, Science Direct, Nature, AIP, Wiley, IEEE explorer, IoP science, ACS publications, and MDPI. Our search also included non-pre-reviewed articles from arXiv. Thus, we performed the critical appraisal using the AACODS (Authority, Accuracy, Coverage, Objectivity, Date, Significance) checklist as an evaluation and critical appraisal tool of grey literature (publications and research created by groups not affiliated with conventional academic or commercial publishing institutions).

We begin with queering all the repositories with different research items. We defined the keywords such as "Machine Learning", "Data-driven approach", "PV cell architecture", "Solar cells", "Low-cost", "Optimization" and "fabrication" shown in table 1 for collecting our research articles. In figure 2 we demonstrate the PRISMA (Preferred Reporting Items for Systematic Review and Meta-Analysis) model showing a screening of the shortlisted publications depending on our research questions. Articles were scanned based on their title and abstract as well as a full-text read of the publications. In addition, we developed search strings using Boolean operators (AND, OR) to connect these keywords.

\begin{figure*}[ht]
 \centering
 \includegraphics[width=18cm]{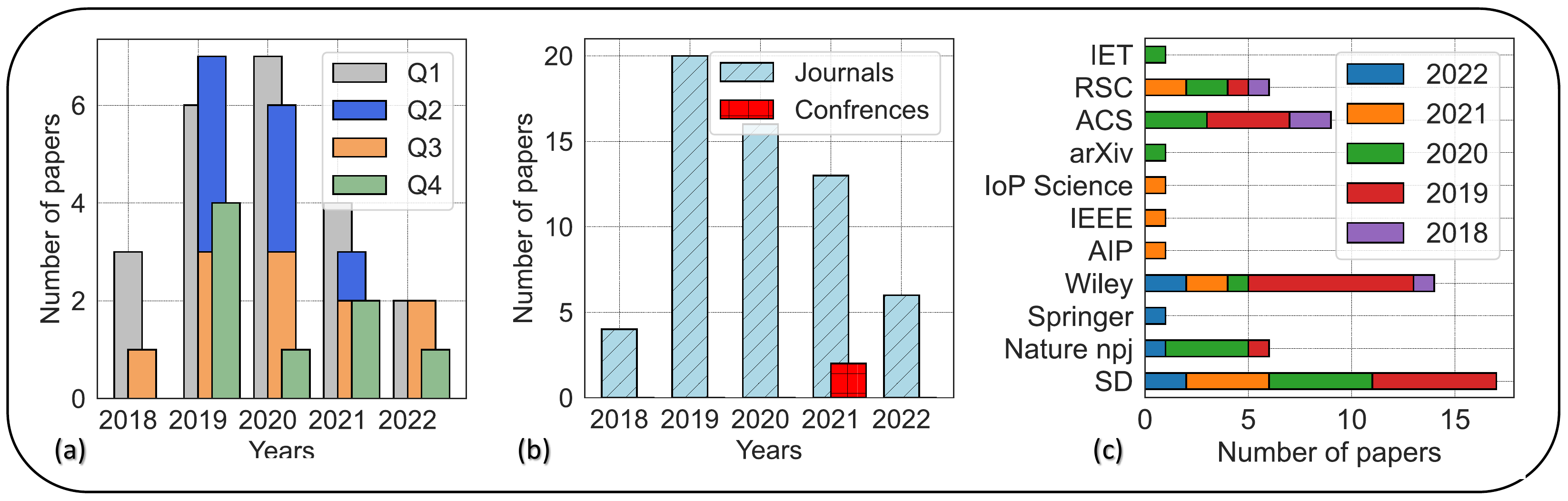}
 \caption{\emph{The figure demonstrates the publication trends for the defined research questions, (a) Number of papers shortlisted as per the research questions from 2018 to 2022., (b) Numerical count of research articles published in the conference or journal consecutively from 2018 to 2022 according to our shortlisted questions., (c) Periodic distribution of achieved articles, research articles and peer-reviewed publications shortlisted depending upon our research questions and research objectives according to the different publishers from 2018 to 2022.  }} 
 \label{fig:PublicationTrends}
\end{figure*}

\subsubsection{Inclusion criteria}
The following are the parameters used in the inclusion criteria.

\begin{enumerate}
    \item We included only English-language articles involving the data-driven approaches of designing solar cells using ML techniques and were pertinent to the study issues such as poor data quantity and data quality.
    \item We included the pertinent articles facilitating the discovery of only low-cost solar cells using ML methods before determining their eligibility.
   \item We included comparative studies involving the optimization and robustness of solar cells designed from ML services.
    \item We targeted only articles that discussed ML for solar cells, solar cell optimization, and publications on ML integration on solar cells.
\end{enumerate}

\subsubsection{Exclusion criteria}

The following is a list of the exclusion criteria for shortlisting the research papers based on our research objectives and targeted research questions.

\begin{enumerate}
    \item Research articles published in languages other than English. 
    \item Research papers that are not available in full text. 
    \item Editorials, survey reviews, abstracts, and brief papers involving secondary studies are excluded.
    \item Articles that did not address the integration of ML approaches with solar cells and the ones that involved the expensive manufacturing of solar cells.
    \item The research articles published before 2018 were also excluded due to the unavailability of quality input data that resulted in poor implementation of ML techniques. 
\end{enumerate}

\subsubsection{Screening phase}

Articles were further screened in two phases. In the first phase, we examined the title and the abstract of each research article to check whether they satisfied our inclusion criteria. In the second phase, we further shortlisted our articles based on their full text. It is worth mentioning that the same piece of writing frequently appeared in various publications. For example, conference papers frequently appear in journals. We take into account the original writing each item was reviewed throughout the screening stage two. At least two of the contributors of this paper who were entrusted with classifying the items as either pertinent or not pertinent might require more research, as finalized until any such item is either published or the authors have a discussion tagged as relevant or not. Survey and review papers were excluded from our review. Finally, each article was carefully classified and evaluated thematically. 

%\section{Review Results}
\subsection{Review Results}
In this section, we discussed the results that we obtained from shortlisting the research articles. The publication trends such as the number of articles published over a period of 5 years, the number of articles per research question, and publishing houses are discussed in detail in this section. In addition, we presented a new state-of-the-art of approach to validate the research articles using the IBM Watson Studio. 

\begin{figure*}[ht]
 \centering
 \includegraphics[width=16cm]{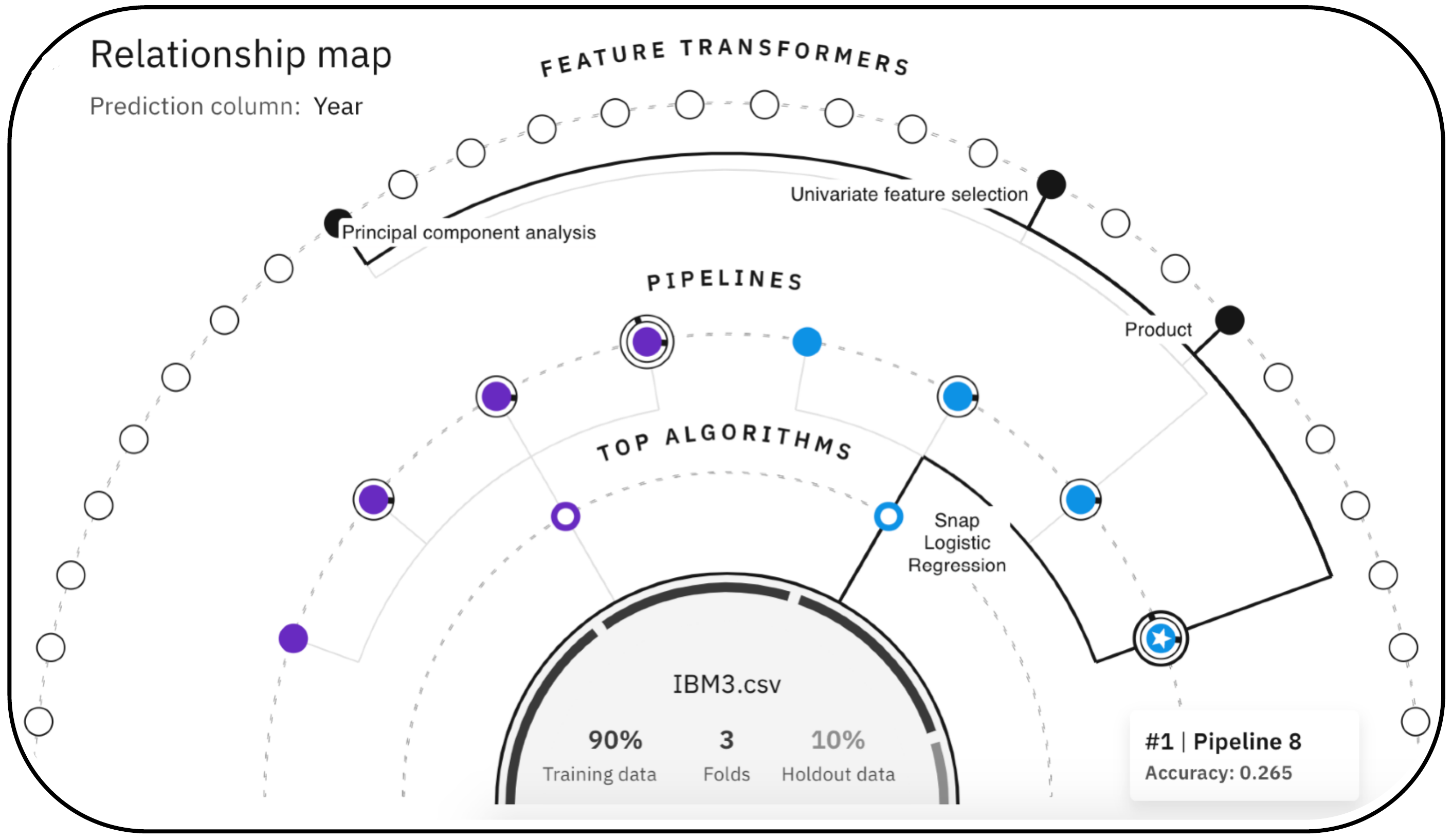}
 \caption{\emph{The figure shows the relationship map of the prediction of the number of research articles published each year using the IBM Watson Studio tool. Also, the figure highlights some insights into the feature transformers, pipelines, and the top ML algorithms involved in the validation of the shortlisted research papers. The input data is provided to the IBM Watson tool using the manually shortlisted research articles from the Google Scholar and Web of Science and accordingly, the AutoAI experiment tool provides the information regarding the research articles published based on our defined research questions for the study under consideration.}}
 \label{fig:Ref3}
\end{figure*}

\subsubsection{Publication trends}

Based on the information presented in the title and abstract, we screened 82 manuscripts that satisfied our search criteria. Following a second screening phase, only 58 papers were relevant to our inclusion criteria. 

In terms of publication trends, it appears that the majority of research articles (67\%) were focused on addressing research questions RQ1 and RQ2, as demonstrated from Figure \ref{fig:PublicationTrends}a. Moreover, only 2 articles were published in IEEE Xplore conference proceedings, as shown in figure \ref{fig:PublicationTrends}b. Consequently, figure 3 (c) represents the bar chart of the distribution of selected publications according to their types for each year. Based on our analysis, we can fairly comment that the maximum number of papers are published in Science Direct in the year 2019, followed by Wiley in 2019 however, the least number of articles are published in IET, IOP Sciences, IEEE Xplore Conferences, AIP, and Springer. Furthermore, most articles were published with Science Direct, Wiley, and ACS, as demonstrated in figure \ref{fig:PublicationTrends}c.

\subsubsection{Validation of Papers}

Further, in order to validate the shortlisted research papers, we used IBM's Watson Studio tool which involves the process of experimentation to deployment, as well as data exploration, model development, and training. IBM Watson Studio is a data science IDE tool designed to help data scientists develop ML models. Moreover, using Watson Studio's 'smart suggestions' we simplified the shortlisted papers from predictions and push models with the Watson ML platform across any cloud.

\begin{figure*}[ht]
 \centering
 \includegraphics[width=18cm]{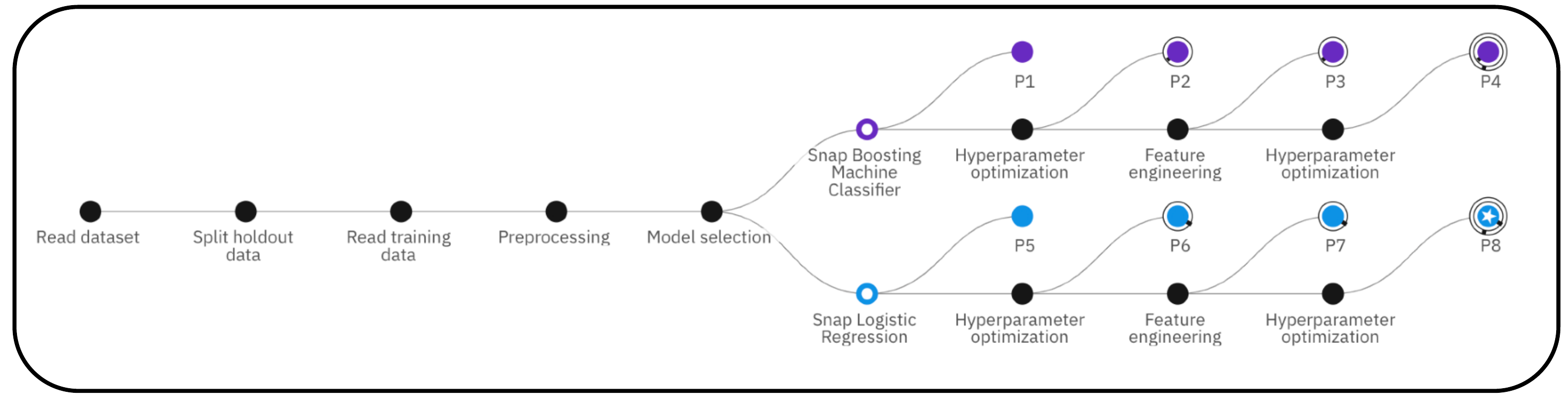}
 \caption{\emph{The pipeline representation of the ML algorithms used to validate the shortlisted papers using the IBM Watson studio tool. Accordingly, the algorithm uses ML techniques such as Snap Logistic Regression, Hyperparameter optimization, feature engineering, and another hyperparameter optimization to determine the most optimized algorithm for predicting the shortlisted research papers.}}
 \label{fig:Ref4}
\end{figure*}

To perform this validation, details of each shortlisted paper were first tabulated in a spreadsheet. Parameters such as the year of the publication, the first author, the publisher, type of manuscript (journal or conference) were fed as input data to Watson Studio's Auto AI tool. Further, to validate the papers, we run a new project under the AutoAI experiment which allows the user to build a fully automated ML model to predict or forecast the parameter under consideration. However, we need to associate different ML or Natural language Programming services and compute the configuration of 8 vCPU and 32 GB RAM. Followed by, once the configuration to run an ML model was set, then we uploaded our data file (IBM3.csv) to our Watson Studio project. Uploading the dataset also gives us an opportunity to visualize the dataset in the form of charts and the Watson Studio tool automatically arranges the dataset to avoid any null values in the data. Therefore, once the input is provided to the Watson Studio model, it processes for the ML algorithm automatically. Thus, we set the predicted parameter i.e. the output result under consideration to be the year of the publication.

In the AutoAI experiment, the tool automatically uses various ML techniques after the analysis of the data. Here, for our model AutoAI applied Multiple Classification prediction types and the model was optimized for RMSE and run time. After the experiment is run on the Watson Studio tool, the dataset is read, split holdout (10\%), read training data (90\%), preprocessing and model selection are performed. Consequently, the relationship map presented in figure 4 describes the best feature transformers, pipelines used, and the top algorithms. Accordingly, the progress map in figure 5 shows the selected algorithm, hyperparameter optimization, feature engineering, and the most optimized feature transformers. Additionally, the most optimized ML model used was Snap Logistic Regression, having pipeline 8 showing the accuracy of the shortlisted papers and lastly, presenting the feature transformers such as the Principal component analysis, Univariate feature selection, and the product. There is a slight discrepancy in the accuracy of the model due to the fact, that the research articles highlighted in different search engines, such as Google Scholar, Web of Sciences, IEEE Xplore, etc., display research articles that are out of the scope of our defined research questions in the methodology section. Also, most of the research articles are repeated at different search engines and whilst doing our manual search of the research paper; we subtracted those articles. 

\section{Results and Analysis}

In this section, we discuss our shortlisted research articles and how they are aligned to our research objectives and questions.  Figure 6 shows the workflow of the planning (data extraction and data pre-processing), training (applying various ML techniques and comparing the model's accuracy), testing (optimization), and execution (fabricating solar cells in the laboratory) for discovering new solar cell architectures. As previously mentioned, our review focuses on low-cost solar cells such as PSCs, OSCs, and hybrids.

\begin{figure*}[ht]
 \centering
 \includegraphics[width=18cm]{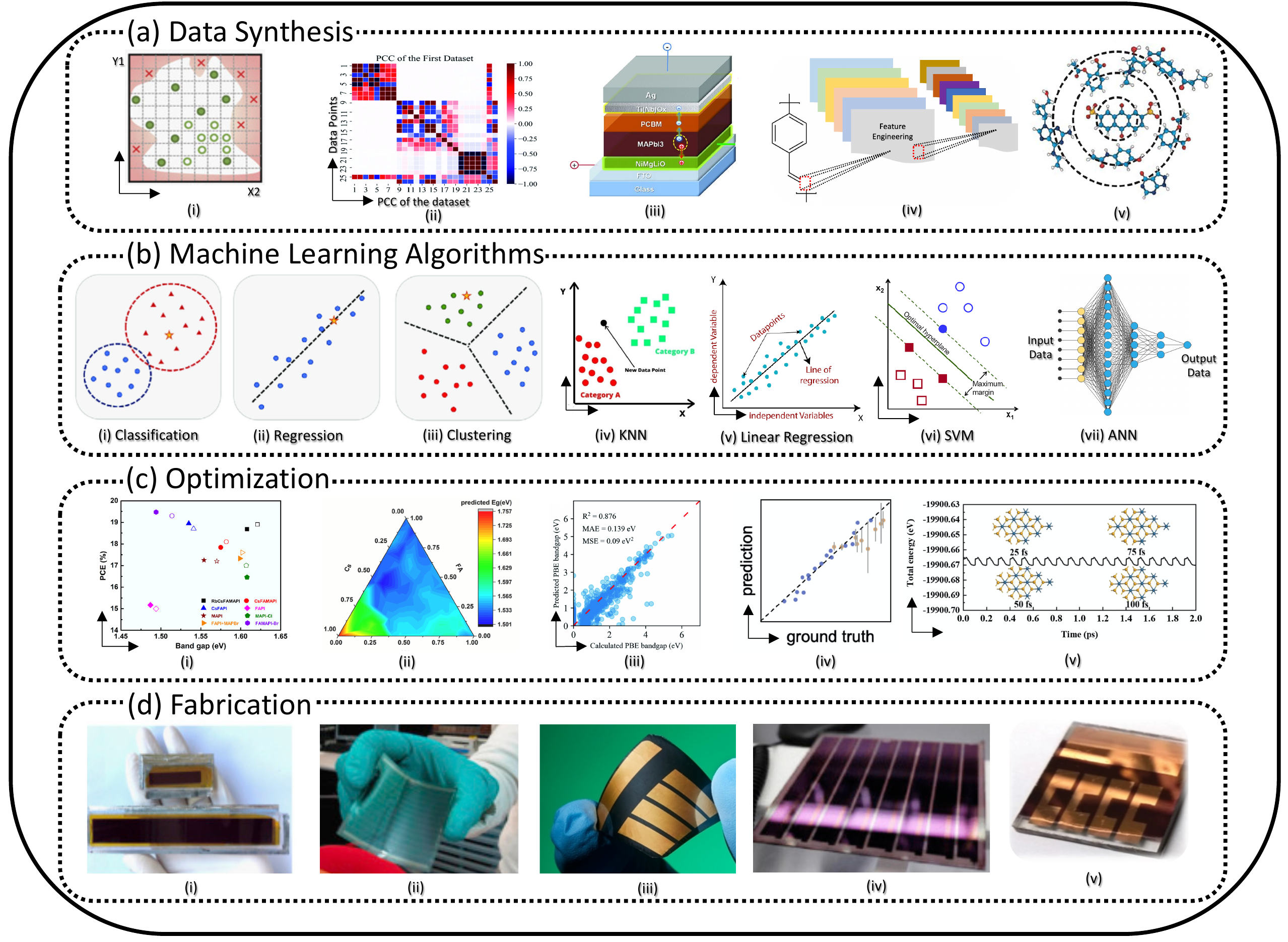}
 \caption{\emph{The figure demonstrates the general workflow of the process of discovering low-cost solar cells using ML algorithms. The block diagram is divided into four block diagrams, (a) data synthesis, (b) ML algorithms, (c) optimization, and (d) fabrication. For (a) data synthesis, (i) Discusses the data extraction in a statistical form, (ii) Pearson's correlation coefficient matrix, (iii) Solar cell architecture with layer combinations, \cite{chen2015efficient},  (iv) data-preprocessing for classification problems \cite{mahmood2021machine} and (v) Gradient-based extraction of data. \cite{wei2019machine} The second block (b) represents the ML algorithms used, (i) Classification, (ii) Regression, (iii) Clustering, \cite{peng2021machine} (iv) KNN, (v) Linear regression, (vi) SVM, (vii) ANN. \cite{hashemi2022machine} The third block (c) discusses the optimization techniques, (i) Bandgap Vs PCE curve, \cite{gok2022predicting} (ii) Ternary contour plots, \cite{liu2022study} (iii) Predicted Vs Calculated PCE, (iv) Predicted Vs Ground truth curve, (v) Predicted accuracy of ML model, (vi) Total energy dissipation Vs Time curve. \cite{zhang2021prediction} The fourth block discusses the fabricated solar cells. (vi) \cite{mathews2019self} \cite{yoo2019interface} \cite{mathews2019technology}}}
 \label{fig:Ref5}
\end{figure*}

\subsection{Data driven approaches for designing low-cost solar cells (Q1)}

Generally, solar cells are designed for achieving certain targets in terms of reliability, affordability, efficiency, and stability. Moreover, in order to forecast the structure of low-cost solar cells, research is in progress to collect and analyze the data generated from previous experiences of solar cell fabrication in a real environment. The quantity and quality of the extracted dataset play a vital role in the effectiveness of ML algorithms. From the literature, more input data leads obviously to higher accuracy and lower functional error values. 

\subsubsection{Perovskite solar cells}

Jino \textit{\textit{et al.}} \cite{im2019identifying} investigated how the Gradient Boost Regression Trees (GBRT) ML method \cite{persson2017multi} can be used for designing Pb-free perovskites. They developed a dataset containing the electronic structures of candidate halide double perovskite. Using the dataset, the GBRT ML model was implemented to predict the values of heat formation and bandgap. Initially, they generated the dataset using two space groups of the crystal structure with 540 hypothetical chemical compounds of $A_{2} B^{1+} B^{3+} X_{6}$. Finally, they conducted statistical analysis on the attributes that were chosen to determine design principles for the development of fresh lead-free perovskites.

Moreover, a study presented by Jinxin \textit{\textit{et al.}} \cite{li2019predictions} showed how 333 data points from nearly 2000 peer-reviewed papers were used to build ML models for designing PSCs. Their ML models included Linear Regression, KNN, RF and Artificial Neural Networks (ANN) for building two forecasting models, material property characteristics and device performance prediction. The higher R-value proves that the expected trend is consistent with actual experiments and PSC physics. The highest theoretically computed solar cell efficiency curve depending on the solar spectrum has a bandgap area in the range of 1.15-1.35 eV, and this bandgap region predicts a PCE of above 25\%.

Moreover, Felipe \textit{\textit{et al.}}, \cite{oviedo2020bridging} demonstrated a new data-driven optimization framework to bridge the mismatch between R\&D and industrial production of solar cells. Further, their framework incorporated scalable inference and techno-economic analysis using ML approaches to predict the root cause of the underperformance in PSCs. They also compared  traditional R\&D optimization vs their proposed total revenue optimization framework using linear, binned and non-linear functions. Consequently, they presented a case study for fabricating 144 PSCs choosing 12 various combinations of dominant processes. In addition, they proposed a surrogate-based black-box model such as Gaussian Process Regression (GPR) and Bayesian Optimization (BO). \cite{nikolaidis2021gaussian} 

In a conference, Maniell \textit{\textit{et al.}} \cite{maniell9325629machine} demonstrated how the optoelectronics properties of PSCs can be predicted using ML methods. A model was developed for testing the bandgap of new different types of PSCs, and the bandgap was capable of predicting the chemical properties and material composition. $C_{Sx}MA_{1-x}PbI_{3}$, $CsPb(I_{x}Br_{1-x})_{3}$ and $MAPb_{1-x}Sn_{x}I_{3}$ were the perovskite materials used for testing and resulted in bandgaps ranging from 1.3-2.3 eV. In addition, their study presented a curve showing the predicted PCE values from the ML model vs the actual PCE from fabricated samples. Moreover, another result showed that the predicted value of the fabricated $CsSnI_{3}$ was 1.15 eV whereas the fabricated sample had a bandgap of 1.25 eV. Lastly, their research article discussed various ML models such as ANN, Random forest algorithm, and Support Vector Regression. 

In addition, the robot accelerated discovery and investigation of PSCs were demonstrated by Zhi Li \textit{\textit{et al.}} \cite{li2020robot}. The article presented an automated, high-throughput method for evaluating single crystals of metal halide perovskites based on inverse temperature crystallization (ITC) in order to quickly pinpoint and perfect the conditions for the synthesis of high-quality single crystals. Using 45 organic ammonium cations, a total of 8172 metal halide perovskite synthesis processes were carried out. The screening enhanced the number of metal halide perovskite materials by five times and resulted in designing a new combination of PSCs such as $[C_{2}H_{7}N_{2}][PbI_{3}]$ and $[C_{7}H_{16}N_{2}][PbI_{4}]$. In addition, to enable experiment generation and data management, they used a software pipeline called ESCALATE (Experiment Specification, Capture and Laboratory Autonomous Technology). Further, their research added 17 new materials (a 400\% increase) of metal halide perovskites, which are accessible via ITC. This helped identify conditions that lead to the formation of perovskite single crystals consisting of 19 of 45 target perovskite compositions.

In 2020, Yun \textit{\textit{et al.}} \cite{zhang2020machine} investigated the ML lattice constants for cubic perovskite $A_{2}XY_{6}$ compounds. Their dataset included a broad spectrum of Fmm group perovskite halides and a total of 79 samples. With lattice constants ranging from 8.109 A to 11.790 A, 79 cubic perovskite compounds were investigated. The ionic radii of [K, Cs, Rb, Tl], [Ge, Mn, Ni, Pd, Pt, Si, Cr, Pd, Ir, Mo, Pb, Re, Se, Ta, Sn, Te, Ti, W, Zr, Ru, Tc, Po, U, Os, Hf], and [F, Cl, Br, I] were among those used as descriptors. The GPR was used for determining the relation between the ionic radii and the lattice constants for cubic perovskites. They used MATLAB for the computational exploration of the model and achieved CC, RMSE, and MAE of 99.72\%, 65\%, and 0.44\%, respectively.

In addition, Chenglong \textit{\textit{et al.}} \cite{she2021machine} presented a two-step ML approach for PSC design, which was based on 2006 PSCs data points taken from peer-reviewed articles published between 2013 and 2020. The authors developed heuristics for high-efficiency PSC and thus, improving PCE dependent on doping of the ETL. The main characteristic of their study was to determine the development of high-performance PCE of PSCs. Their research showed that using $SnO_{2}$ and $TiO_{2}$ ETLs, mixed-cations perovskites, dimethyl sulfoxide, and dimethylformamide, as well as anti-solvent treatment, led to even higher PCEs. Lastly, they predicted that FA-MA-based PSC with a Cs-doped $TiO_{2}$ ETL and a Cs-FA-MA-based PSC with an S-doped $SnO_{2}$ ETL were also expected to show PCEs of up to 30.47\% and 28.54\%.

To expedite the identification of prospective PV cells from 2D perovskites, Hong-Jian \textit{\textit{et al.}} \cite{feng2021machine} integrated atomic-level prediction with ML and DFT. Their model implemented a gradient boosting regressor (GBR), a random forest regressor (RF), and an extra tree regressor (EXTR) ML for training  a dataset of 2303 perovskite materials. Further, the trained model screened out 4828 materials and also pre-screened using DFT structural relaxation validation from 29,285 artificial perovskites. In fact, a maximum PCE of 30.35\% and 26.03\% was achieved for ($Sr_{2}VON_{3}$ and $Ba_{2}VON_{3}$).

Likewise, Elif \textit{\textit{et al.}} \cite{gok2022predicting} predicted the overall performance and bandgap in PSCs. In her analysis, she used eight different PSCs to forecast the bandgap and PCE of perovskites. Initially, they performed the bandgap estimation of perovskites from Tauc plots on a UV-vis spectroscopy using the RF regression ML model with more than one decision tree and experimental approach. Later, they developed a model showing the J-V spectra predicted values for calculating the PCE. Their results showed that perovskites with bandgaps exceeding 0.99 eV could be used to model various new lead halide structure perovskites depending on the accurately predicted value of the bandgap.

Another case study presented by Xia \textit{\textit{et al.}} \cite{cai2022data} combined ML techniques with an efficient forward-inverse method to research $MAS_{nx}Pb1_{x}I_{3}$ material and explored high-performance PSCs. With 14 physicochemical parameters and the Sn-Pb ratio as inputs, the $E_g$ model of $MAS_{nx}Pb1_{x}I_{3}$ was first developed for forward analysis, and the asymmetrically bowing relationship between the Sn-Pb ratio and the Eg of OMHP was used. The established NN-based models for PSC performance models showed good predictions for the data points and offered significant insights for PSC devices. Further, for the performance model, a comparison of the prediction model was made with the ML algorithms such as LR, SVR, KNR, RFR, and GBR. In fact, ML models with GBR performed best with values of R2, RMSE, and MAE reaching 0.9172, 0.0386, and 0.0325. 

\begin{figure*}[ht]
 \centering
 \includegraphics[width=18cm]{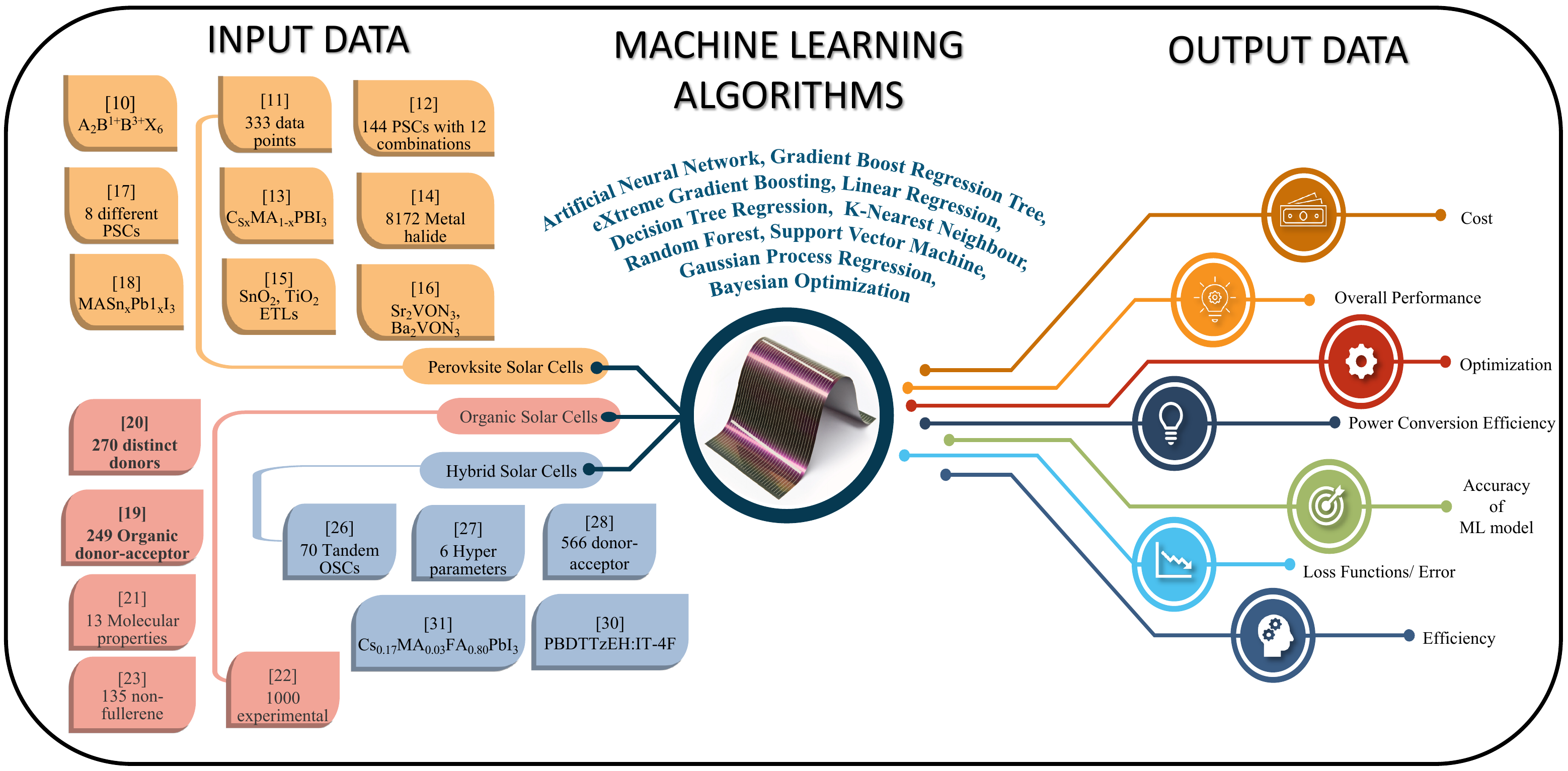}
 \caption{\emph{The figure provides information on input data for various materials that are reviewed based on our defined research questions for 3 types of solar cells such as PSCs, OSCs, and hybrid. The majority of ML algorithms used in the process are highlighted to determine the resultant output in terms of electrical characteristics of re-configurable solar cells. The numbers in the box of the input data section are linked with reference numbers answering our RQ1 for PSCs, OSCs, and hybrid solar cells.}}
 \label{fig:Ref6}
\end{figure*}

\subsubsection{Organic solar cells}  

A rigorous framework involving the classification of the chemical structures in materials discovery was presented by Shinji \textit{et al.}\cite{nagasawa2018computer}. Further, the dataset of 249 Organic donor-acceptor pairs was computed based on equilibrium geometries and electronic properties such as DFT simulations. Initially, their study discussed predictions using Scharbar's model and resulted in a small energy bandgap of 1.5 eV between the experimental and the computational energy bands. Moreover, they implemented k-NN regression for predicting OSCs characteristics and their PCEs. Finally, the study concluded that k-NN results in correlations of 0.6, which were further improved to 0.7 by implementing non-linear kernel methods.

In addition, Harikrishna \textit{\textit{et al.}} \cite{sahu2018toward} investigated the PCE of OSCs using ML techniques. They developed a dataset of 280 small molecule OSCs with 270 distinct donors. Firstly, they analyzed the significance of orbitals in the energy conversion process and developed ML models using the characteristics of organic compounds to estimate the PCE for high throughput virtual screening. In another study, they implemented ML methods to study the correlations between the molecular properties and the device characteristics of an OSC \cite{sahu2019unraveling}. The authors designed ML methods based on 13 molecular properties as descriptors to predict the three device parameters such as ($V_{OC}$, $J_{SC}$, and the fill factor). In addition, the calculations were carried out on Gaussian 09 package for a computational server having Intel Xeon 5115 CPUs. They combined multiple regression trees along with RF and GBRT to incorporate the ML methods. Further, screening of the potential compounds by these models results in high predictive ability (r = 0.7).

Moreover, Daniele \textit{\textit{et al.}} \cite{padula2019combining} performed the computer-aided screening of polymers-based OSCs using RF and ANN-based ML supervised-learning models. The dataset involved 1000 experimental characteristics such as PCE, the molecular weight of each organic compound, and other electronic properties. The results showed that the correlation coefficient of ANN was low. However, the RF model achieved better accuracy than the predictive model. Subsequently, Min-Hsuan \textit{\textit{et al.}} \cite{lee2020robust} also performed the RFT regression for the analysis of the non-fullerene-based OSCs to predict the overall efficiency of the solar cells. A dataset of 135 non-fullerene acceptor/donor pairs based on OSCs (117 non-fullerene acceptor materials and 30 donor materials) was gathered to examine its electronic properties and device performances. Therefore, their ML model resulted in the highest predictive power by achieving the coefficient of determination ($R^{2}$) of 0.85 for the training and 0.80 for testing sets of the ML algorithm.

Furthermore, Xiaoyan \textit{\textit{et al.}} \cite{du2021elucidating} demonstrated an optimization technique to assess the potential of organic photovoltaic (OPV) materials and solar cell devices for industrial production. They presented an automated characterization of OPV materials, device performance and photostability. The GPR ML technique drove the optimization method with optical absorption characteristics and indicated better prediction accuracies for PV electrical characteristics. Moreover, the efficiency and photostability screening for 100 process conditions were completed in 70 hours. They also proposed a model material system of PM6:Y6, completely automated device fabrication in air resulted in a maximum PCE of 14\%.

In one of the latest papers published by Ahmad \textit{\textit{et al.}} \cite{irfan2022learning}, they discuss the implementation of ML to screen small molecule donors for OSCs and molecular descriptors feed ML methods. The co-authors collected a dataset of 340 OSCs devices with donors represented as small molecules while acceptors as fullerenes for the ML-assisted pipeline suitable for small molecule donors for Y6 (an electron acceptor). In addition, they performed ML analysis on an open-source platform called Konstanz Information Miner (KNIME). Further, for training the model, the dataset was divided into training sets, validating sets and external test sets. Also, the descriptors and experimental PCE were used as input to the ML model. They compared the result depending on various regression techniques, such as RF, LR, SVM and k-NN, for the prediction of PCE. Using data from small donors paired with fullerenes, the SVM model was trained and showed higher prediction ability. The PCE of a few small molecule donors linked with Y6 was predicted using their approach and developed are more than 1000 new small molecule donors. Accordingly, the PCEs were anticipated, and the top 10 applicants with a PCE of over 13\% were chosen in their study.

\subsubsection{Hybrid Solar Cells}

Another article presented by Min-Hsuan \textit{\textit{et al.}} \cite{lee2020performance} investigated the performance and matching band structure for Tandem OSCs by implementing two ML methods, RF and the SVR. The ML techniques were initially developed using 70 tandem OSCs (37 conventional and 33 inverted tandem OSCs), which were used as the data points. Furthermore, to understand the structure, they calculated Pearson's correlation coefficient. Among the two ML methods, the efficient method for forecasting solar efficiency was the RF Regression having eight electronic features of selection.

Moreover, to address the stability concerns with PSCs, Tianmin \textit{\textit{et al.}} \cite{wu2020deep} used a progressive ML algorithm to investigate the impact of input data by providing a reliable and accurate approach for deep mining of the hidden hybrid organic-inorganic solar cells. To predict the electronic bandgaps of HOIP perovskites, they implemented GBR, SVR, and kernel ridge regression (KRR) using material property. The best results from six hyperparameters were chosen. They also used DFT calculations for the chosen HIO perovskites and incorporated them into the Vienna Ab-initio simulation package (VASP). Their results show that the GBR model performs with the highest level of accuracy (R2 = 0.943, MAE = 0.203, MSE = 0.086) when compared to the SVR (R2 = 0.826, MAE = 0.367, MSE = 0.276) and KRR (R2 = 0.819, MAE = 0.387, MSE = 0.288) models.

The effect of enhancing the descriptors using ML prediction for small molecule-based OSCs was discussed by Zhi-Wen \textit{\textit{et al.}} in his study. \cite{zhao2020effect} The dataset consists of a total of 566 organic donor-acceptor (D/A) pairs found from the literature search, with 513 unique donors and 33 unique acceptors (including $C_{60}$, $PC_{61}BM$, $PC_{71}BM$, ITIC, IDTBR, IDIC, PDIs, etc.) among the donors. Further, they implemented k-NN, KRR and SVR ML models to predict the PCE of hybrid solar cells. Also, the study examined Pearson’s correlation coefficient for all combinations of descriptors, including donor molecules and device parameters.

In another study presented by Yao \textit{\textit{et al.}}, \cite{wu2020machine} five different ML algorithms were used and gave 565 donor-acceptor combinations for training the dataset. Furthermore, to implement the material design and donor-acceptor pairs, the screening of non-fullerene in OSCs was performed. They used 565 donor/acceptor (D/A) combinations as training data sets in their study to assess the viability of these ML algorithms for use in directing material design and the screening of D/A pairs. Therefore, the ML techniques RF and BRT offer the best prediction capacities. Additionally, RF and BRT models are screened and estimated to be more than 32 million D/A pairs, respectively. Lastly, six photovoltaic D/A couples are picked and synthesized so that their experimental and predicted PCEs for critical comparison.

In an investigation presented by Kakaraparthi \textit{\textit{et al.}} \cite{kranthiraja2021experiment}, the co-authors used the RF model on an experimental dataset consisting of 0.85 correlation coefficient for the ML of non-fullerene and polymer OSCs. Moreover, 200,932 conjugated polymers produced by the combinatorial coupling of acceptor and donor units were screened virtually. Additionally, a number of conjugated polymers centred on benzodithiophene and thiazolothiazole were created, produced, and studied using various alkyl chains in order to assess the efficacy of the ML model. In terms of the selection of alkyl chains, PBDTTzEH: IT-4F demonstrated a PCE of 10.10\% and, thus, shows good predictions while using ML techniques.

One of the primary concerns with perovskites is their stability. As a result, Shijing \textit{\textit{et al.}} \cite{sun2021data} demonstrated how to discover the most stable organic-inorganic alloyed perovskites using a sequential learning framework. They introduced a data-fusion approach for estimating Gibbs Free Energy of mixing from DFT and experimentally analyzed degradation using aging tests. Moreover, they applied ML probabilistic constraints in an end-to-end BO approach to combine data from high-throughput degradation testing and first-principle simulations of phase thermodynamics. The results showed that perovskites centered at $Cs_{0.17}MA_{0.03}FA_{0.80}PbI_{3}$ exhibit low optical change with increased temperature, moisture, and light having more than17-fold stability improvement over $MAPbI_{3}$ by sampling 1.8\% of the discretized $Cs_{x}MA_{y}FA_{1xy}PbI_{3}$ compositional space ($MA$, methylammonium; $FA$, formamidinium; $PbI_{3}$, lead halide).

\subsubsection{Natural Language Process}

In another study, a framework related to the high-throughput synthesis of the PSCs was discussed with ML image recognition used for automated characterization by Jeffrey \textit{\textit{et al.}} \cite{kirman2020machine}. Perovskite single-crystal synthesis was carried out at high throughput, and the results were identified using convolutional neural network-based image recognition. Also, they quickly created 96 distinct crystallization environments using a protein drop setter and then examined the crystals. On the other hand, trained a convolutional neural network (CNN) was used to determine if crystals had been produced using a dataset of 7,000 photographs. Then, a larger dataset of 25,000 photos was employed with this classifier. The first synthesis of $(3-PLA)_{2}PbCl_{4}$ was then achieved after they employed ML modeling to predict the ideal conditions for synthesizing a novel perovskite single crystal.

A study presented by Lei \textit{\textit{et al.}} \cite{zhang2022unsupervised} showed ML techniques based on natural language processing (NLP) to predict the properties of solar cell materials, which were then examined using first-principle calculations. The aim of the study was to reduce the amount of human interaction and enable computers (without supervision) to learn the latent knowledge about solar cell materials depending on the textual data and generate predictions about the composition of solar cells. The first-principles calculations were used to determine the projected material's density of states, UV–vis absorption spectra, as well as band structures in order to assess their suitability for photovoltaic applications. The formula and targeted keywords for solar cells were represented as vectors in the ML process, which facilitated the successful relationship extraction of the materials and their applications. The  ML model was validated using first-principles calculations on the unusual solar cell materials included in the list, and the projected candidates, such as $As_2O_5$ have good electrical and optical characteristics that are suitable for solar cell applications.

%-------------------------------------------------------------------------------------------------------------------------------------------------------------------------

% Targeting Question 2 of our research question 
\subsection{ML to Facilitate the Discovery of Solar Cells (Q2)}

This section discusses the research articles and peer-reviewed journals related to the discovery of solar cells using ML techniques.

\subsubsection{Discovery of Organic Structures}

A target-driven approach was provided by Tianmin \textit{\textit{et al.}} \cite{wu2019global} to accelerate the discovery of HOIPs for PV applications from 230808 HOIP candidates. Also, they combined the ML method with DFT calculations. 686 orthorhombic-like HOIPs with the appropriate bandgap were chosen after possible HOIP candidates are subjected to the two criteria of charge neutrality condition and stability condition, followed by an ML screening. In ML screening, ensemble learning was used to forecast the bandgap of 38086 HOIPs candidates using three ML models, including GBR, SVR, and KRR. Finally, 132 stable and non-toxic orthorhombic-like HOIPs (free of Cd, Pb, and Hg) were confirmed by DFT calculations with the proper band gap for solar cells.

Oleksandr \textit{\textit{et al.}} \cite{voznyy2019machine} used ML in-the-loop to learn from the experimental data, suggested experimental parameters to explore, and indicated regions of synthetic parameter space that would permit record-monodispersity PbS quantum dots. Their results show that the technique that produces record-large bandgap (611 nm exciton) PbS nanoparticles with a well-defined excitonic absorption peak (half-width at half-maximum (hwhm) of 145 meV) permits nucleation to triumph overgrowth by adding a growth-slowing precursor (oleylamine). With a hwhm of 55 meV at 950 nm and 24 meV at 1500 nm, respectively, as opposed to the best-published values of 75 and 26 meV, they also improved monodispersity at longer wavelengths.

\begin{figure*}[ht]
 \centering
 \includegraphics[width=18cm]{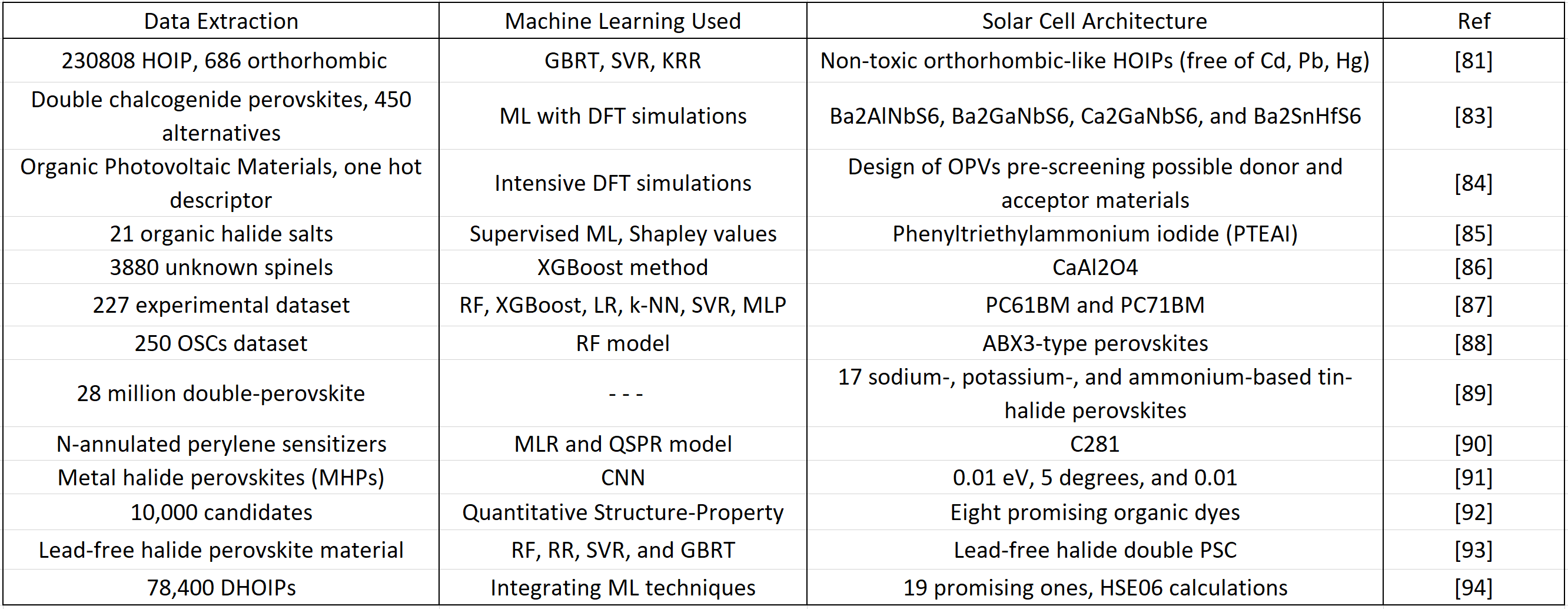}
 \caption{\emph{Literature Discussing the ML for facilitating the discovery of solar cells.}}
 \label{fig:Ref16}
\end{figure*}

Double chalcogenide perovskites were investigated in a study presented by Michael \textit{\textit{et al.}} \cite{l2019machine} to find new photovoltaic absorbers that can take the place of CH3NH3PbI3. ML approaches were used to categorize materials as potential photovoltaic absorbers using information from the periodic table, thus avoiding unnecessary computation due to the wide range of possible compounds. On the created data set, a random forest method obtains a cross-validation accuracy of 86.4\%. Traditional and statistical approaches are used to identify over 450 potential alternatives, with Ba2AlNbS6, Ba2GaNbS6, Ca2GaNbS6, Sr2InNbS6, and Ba2SnHfS6 emerging as the most promising options when thermodynamic stability, kinetic stability, and optical absorption are taken into account.

Nastaran \textit{\textit{et al.}} \cite{meftahi2020machine} in a study showed that ML techniques used by computationally intensive DFT simulations to quickly and precisely estimate the properties of OPV materials. One-hot descriptors, OPV power conversion efficiency (PCE), open circuit potential ($V_{oc}$), short circuit density ($J_{sc}$), highest occupied molecular orbital (HOMO) energy, lowest unoccupied molecular orbital (LUMO) energy, and the HOMO-LUMO gap were all quantified in the study. With a standard error of 0.5 for a percentage of PCE for both the training and test sets, the most reliable and predictive models were able to predict PCE (computed by DFT). Their methodology helps to expedite the design of OPVs for use in green energy applications by pre-screening possible donor and acceptor materials.

An ML framework introduced by Noor \textit{\textit{et al.}} \cite{hartono2020machine} involved optimizing the capping layer of perovskite degradation. They featured 21 organic halide salts, used them as capping layers on (MAPbI3) films, aged them rapidly, and implemented supervised ML and Shapley values to identify factors determining stability. They discovered a correlation between higher MAPbI3 film stability and organic molecules' limited number of hydrogen-bonding donors and tiny topological polar surface area. Phenyltriethylammonium iodide (PTEAI), the best organic halide, successfully increases the stability lifespan of MAPbI3 by 4 2 times over bare MAPbI3 and 1.3 0.3 times over cutting-edge octylammonium bromide (OABr).   

Zhilong \textit{\textit{et al.}} \cite{wang2021accelerated} created a target-driven approach that makes use of ML to speed up the ab initio predictions of unidentified spinels from the periodic table. Eight spinels with direct band gaps and thermal stabilities at room temperature are successfully selected out of 3880 unknown spinels using this method ($CaAl_{2}O_{4}$, $CaGa_{2}O_{4}$, $SnGa_{2}O_{4}$, $CaAl_{2}S_{4}$, $CaGa_{2}S_{4}$, $CaAl_{2}Se_{4}$, $CaGa_{2}Se_{4}$, $CaAl_{2}Te_{4}$). A semiconductor classification model is developed based on the XGBoost method, and it has a strong structure-property link. It has a high prediction accuracy of 91.2\% and a low computational cost of a few milliseconds. The suggested target-driven strategy enables the discovery and design of a wide variety of energy materials while cutting the research cycle of spinel screening by about 3.4 years.

The accuracy for predicting the bandgap of an OSC is a vital factor in terms of the characterization of solar cell devices. Accordingly, Yiming \textit{\textit{et al.}} \cite{liu2022study} used ML algorithms to predict the performance of different architectures for the compound $ABX_{3}$-type in PSCs. Also, they gathered 227 experimental datasets consisting of the bandgap of perovskites extracted from recently published 1254 publications. For their model, they used ML methods such as RF, XGBoost, LR, k-NN, SVR, and Multilayer perceptron (MLP). Their prediction analysis from ML models showed that B-site metal and the X-site halogen ion have a significant impact on bandgaps of the $ABX_{3}$-type perovskites from SHAP explanations. 

Muhammad \textit{\textit{et al.}} \cite{janjua2022machine} did the critical analysis of the small-molecule donors for OSCs such as Fullerene using the ML methods. In order to train the ML model, they used molecular descriptors as an input and consecutively, they implemented a number of ML techniques to measure the best ML algorithm for the desired outcome. The dataset used in the study consists of 250 OSCs having a combination of acceptors and donors as fullerenes ($PC_{61}BM$ and $PC_{71}BM$). They used the platforms like Konstanz Information Miner (KNIME) and Weka platforms to implement the ML model and thus, the Random Forest model resulted the best predictive model with Pearson's coefficient as 0.93. Lastly, to determine the most efficient materials, the PCE values for the small-molecular donor was predicted.

\subsubsection{Discovery of Hybrid Halide Structures}

With multiple newly developed, computationally economical, and high-performing (Pearson's correlation coefficient = 0.7-0.8) ML models employing pertinent descriptors, Harikrishna \textit{\textit{et al.}} \cite{sahu2019designing} carried out high-throughput virtual screening of 10,170 candidate compounds, assembled from 32 distinct building blocks. Furthermore, to create effective molecules, crucial building elements are recognized, and new design principles are implemented. Additionally, 126 candidates are suggested for synthesis and device fabrication with theoretically projected efficiency >8\%.

A high-throughput material search scheme based on materials informatics was devised and carried out for PSC materials after Shohei \textit{et al.} \cite{kanno2019alternative} explored the existence of viable alternative perovskites. More than 28 million double-perovskite-like compounds were screened using this method. Five well-known organic-inorganic tin-halide perovskites and 17 sodium-, potassium-, and ammonium-based tin-halide perovskites were among the 24 most promising possibilities found. Promising solar cell materials included two perovskites based on transition metals.

Further, Lifei \textit{\textit{et al.}} \cite{ju2020accelerated} constructed N-annulated perylene sensitizers and put forth one goal-directed approach that combined quantum chemical analysis with data mining approaches. By using MLR to build the robust quantitative structure-property relationship (QSPR) model, they were able to identify the key characteristics using a genetic algorithm (GA). The potential dyes were then created using the model's recommendations. The proposed molecules' overall power conversion efficiencies (PCEs) were anticipated by the model to be 15.7\%, up 22.0\% from reference dyes $C_{281}$.

For the electrical characteristics of metal halide perovskites (MHPs), which have a billions-range materials design space, Wissam \textit{et al.} \cite{saidi2020machine} employed CNN to create a predictive model. Furthermore, they demonstrated that as compared to simple techniques, a well-designed hierarchical ML strategy offers a higher degree of predictability in terms of MHP features. The bandgap for the MHPs' lattice constants, octahedral angle, and RMSE were all calculated using the hierarchical ML scheme, and the corresponding RMSE values were 0.01 eV, 5 degrees, and 0.01.

Yaping \textit{\textit{et al.}} \cite{wen2020accelerated} combined ML with computational quantum chemistry results in the establishment of an accurate, reliable, and interpretable QSPR model. Using this model, virtual screening as well as the evaluation of synthetic accessibility are carried out to find new effective and synthetically accessible organic dyes for DSSCs. Finally, out of almost 10,000 candidates, eight promising organic dyes with high power conversion efficiency and synthetic accessibility were eliminated.

Moreover, Zongmei \textit{\textit{et al.}} \cite{guo2021machine} investigated the discovery of PSC materials via ML stability and calculated the bandgap of lead-free halide perovskite materials. They performed a comparative analysis of four different ML techniques such as the random forest, ridge regression, support vector regression, and the gradient boost regression tree. Among these four ML techniques, XGBoost gave the highest predictive performance i.e. R2:0.9935 and MAE:0.0126 in terms of thermodynamic stability, and accordingly, the random forest gave the highest predictive performance i.e. R2:0.9410 and MAE:0.1492 for bandgap analysis of the lead-free halide double PSCs. Moreover, their study showed an interesting result that XBoost performs best when considering the thermodynamic stability and electronegativity's linear correlation.  

By integrating ML techniques, high-throughput screening, and density functional theory, Jialu \textit{\textit{et al.}} \cite{chen2022delta} showed the ability to speed up the discovery of double hybrid organic-inorganic perovskites (DHOIPs). In contrast to other studies, the anisotropy of organic cations of DHOIPs was first assessed, and then the properties were predicted using an ML technique using low-level calculations to predict the properties of DHOIPs accurately. From 78,400 DHOIPs, 19 promising ones with suitable bandgaps for solar cells were selected and verified using HSE06 calculations.

John \textit{\textit{et al.}} \cite{howard2019machine} investigated the bias, temperature, light, and H2O, O2, and air pressure affected device performance and recovery. They first talked about important studies that assess the 3R cycle's capabilities of perovskites and how ML algorithms may help determine the best values for each operating parameter. They then looked at perovskite dynamics and degradation, highlighting the difficulties in understanding this 3R cycle. Finally, they suggested an ML paradigm with a shared knowledge library for improving long-term performance and forecasting device performance recovery.

\subsection{ML for Solar Cell Optimization (Q3)}

This section of our systematic review discusses our third research question i.e. the optimization techniques that are applied using ML algorithms for developing reconfigurable and optimized solar cells. The technical research articles that showed experimental work for implementing the ML algorithms for discovering the optimized solar cells are included. 

\begin{figure*}[ht]
 \centering
 \includegraphics[width=16cm]{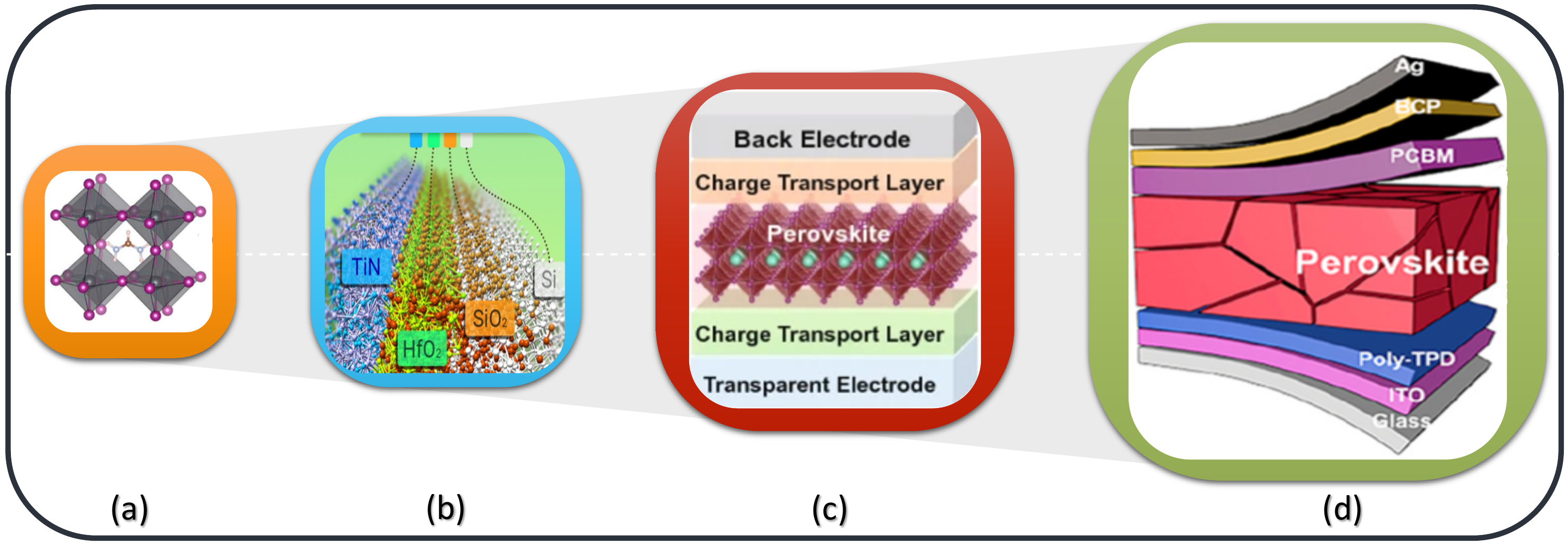}
 \caption{\emph{The figure describes the internal solar cell architecture with multiple layers and the chemical components required in the designing of re-configurable solar cells. In addition, (a) represents the chemical structure of the perovskite with carbon composition, \cite{chan2019high}(b) Depicts the arrangement of the chemical components in a solar cell, (c) shows various layers of the solar cell that is sliced to give a clear image of the solar cell architecture, \cite{wu2021main} and (d) is the outer layer of the solar cell having Ag, BCP, PCBM, Perovskite, Poly-TPD, ITO and glass. \cite{li2020reduced} }}
 \label{fig:Ref7}
\end{figure*}

\subsubsection{Donor/acceptor ratio for higher PCE}
Most scientific advancements in the field of materials have been produced experimentally, frequently using one variable at a time testing. However, neither are the properties of materials-based systems straightforward nor related. Authors in \cite{cao2018optimize}, claim that the optimization of OSCs has a high level of complexity due to the high complexity and interconnectivity of different components. Changing one component can have an unforeseen impact on other components. Hence ML can play a vital role in the optimization process of OSCs. They used $PDCTBT:PC_{71}$ solar cell and observed the effect of donor/acceptor ratio, total concentration, spin speed, and additive volume on PCE(\%). The authors applied SVM using the radial basis function. They conducted two sets of experiments, where they used optimized results of the first experiment in the second experiment and found a significant increase in PCE of fabricated devices. In the first set of the experiment, only three out of fifteen devices were above the threshold (PCE~6.3\%); however, in the second, all thirteen  devices produced PCE above the threshold.

\subsubsection{Conductivity optimization of solar cells}
SVM regression was used in \cite{wei2019machine} for the optimization of $p-CZS/n-si$, $p-CZS/p^+ n-Si$ heterogeneous solar cells. SVM was implemented with a radial-based function using Scikit-learn \cite{pedregosa2011scikit} in python. They used ten-fold cross-validation to tackle the problem of over-fitting. They predicted the figure of merit (FOM) from film conductivity and optical transmission in desired transmission range. Optimization results show that FOM has increased from $14.8 \mu$ to $173 \mu$. Furthermore, current density has increased from 11.8 to 17.9 $mA/cm^2$ for $p-CZS/n-si$ solar cells and from 13.8 to 18.0$mA/cm^2$ $p-CZS/p^+ n-Si$ for solar cell.  The authors claimed their approach is valid for any general application to any material synthesis process with multiple parameters.

\subsubsection{Donor/acceptor material for higher PCE}
From 2010 to 2017, 320 organic donor and acceptor pairs (hetero-junction solar cells) were reported in the literature. These 320 donors and acceptors can make 19912 combinations. Authors in \cite{padula2019concurrent} applied distanced-based ML techniques KNN and SVM to optimize PCE. They have provided a list of unexplored donor and acceptor combinations that can be helpful in the future in fabricating highly efficient solar cells. The use of back propagation neural network, deep neural network, SVM, and the random forest is reported in \cite{sun2019machine} to predict highly efficient OSCs. The data set contained 1719 realistic donor materials of OSCs. The authors used images, ASCII strings, and fingerprints as input, and out that fingerprints with 1000 bits can provide higher conversion efficiency. The authors also proposed ten new materials.  

\subsubsection{Stability optimisation}
Stability is a good indicator of the life span of a solar cell. Multiple parameters can affect the stability of OSCs. Authors in \cite{david2020enhancing} optimized these parameters using sequential minimal optimization regression on a data set obtained from the website of Danish Technical University (DTU) \cite{david2019using}. Authors have presented shortlisted layer-wise materials with the highest weights in sequential minimal optimization regression. These materials are the most influential materials governing the stability and performance of OPV devices.

%A high throughput semi automated experimental platform that uses ML to optimise the electrical conductivity was presented in. IN this study Bayesian optimization was used to sample target data. 

\subsubsection{Copper content optimization in CdTe solar cell}
Cu is essential in CdTe solar cells as back contact and doping agent. Diffusion depth optimization of Cu resulted from diffusion annealing, and cool-down in the fabrication of CdTe solar cell was reported in \cite{salman2021machine}. ANN predicts data generated from software simulation using the Keras library in python. ANN was fed with temperature and duration of diffusing process time. Results show that the predicted and actual depths are only $0.009 \mu$ apart.

\subsubsection{Optimization of diode model for solar cell simulations}
A bio-inspired Modified spotted hyena optimization algorithm was implemented in \cite{gafar2022optimal}, to compare one diode model, two diode modes, and three diode model solar cells in MATLAB. The authors obtained I-V and P-V curves. They found that the three-diode model is the most accurate model.

\subsubsection{Optimisation spray plasma processing}

Optimization is a common theme in materials research when synthesizing a particular material or determining the ideal processing conditions to obtain the desired attribute. The difficulties emerge from the fact that there are several parameters whose weights might influence the outcomes. Additionally, gathering experimental data takes time and money. Authors in \cite{hsu2022accelerate}, presented the work of \cite{gelbart2014bayesian}, where BO was used to optimize the rapid plasma process. The authors used six different parameters are input that affect PCE: linear speed of pray, substrate temperature, the flow rate of precursor, gas flow rate into plasma nozzle, the height of plasma nozzle, and plasma duty cycle, while some other parameters were kept constant such as precursor formulation, concentration, etc. The optimization result showed that PCE increased from 15\% to 17 \%.

%----------------------------------------------------------------------------------------------------------------------------------------------------------------------------

\subsection{ML used for the efficient fabrication of solar cells (Q4)}

The majority of research articles discuss the different types of ML algorithms used to efficiently fabricate the PSCs. In this section, we discuss the most optimum ML algorithms that have proved to find the appropriately efficient technique for fabricating PSCs.

%\subsubsection{ML to design a re-configurable PSC}

PSCs are cheap to fabricate and, as a result, most researchers fabricate these low-cost solar cells by trial and error. Also, fabricating a solar cell consists of a large percentage of permutation and combinations of various physical parameters such as materials used, doping layers, the thickness of the different layers, meshing, contacts, bulkiness, etc. In addition, the solutions-based techniques of fabricating solar cells require less time to manufacture however, exhibits stability concerns. Therefore, we review the ML methods for designing a re-configurable PSC. 

%Remaining papers of Q4 are reviewed in this section

Zhe \textit{et al.} \cite{liu2022machine} demonstrated a sequential learning architecture for producing PSCs that are guided by ML. They applied different methods to create open-air perovskite devices using the rapid spray plasma processing (RSPP) method. Further, showed the best outcome from a device made by RSPP was an efficiency improvement of 18.5\% with a limited experimental budget of screening 100 process scenarios. They achieved this mainly due the three innovations such as flexible knowledge transfer between experimental processes by using prior experimental data as a probabilistic constraint, incorporation of both subjective human observations and ML insights when choosing the next experiments, and adaptive strategy of locating the region of interest using BO before conducting local exploration for high-efficiency devices.

Another research article presented by Vincent \textit{et al.} \cite{le2021identification} discussed a quick and simple tool for identifying the primary losses in PSCs. To comprehend the light intensity dependency of the open-circuit voltage and how it relates to the main recombination mechanism, their model used large-scale drift-diffusion simulations. The ML algorithm was developed using more than 2 million simulations and resulted in a prediction accuracy of up to 82\%.

In addition, Xabier \textit{et al.} \cite{rodriguez2021accelerating} in their study used big data for the discovery of OSCs, such as non-fullerene acceptors and low-bandgap donors-based polymers. Also, they discussed the computational techniques used to choose the most promising chemical molecules from the online material libraries. Secondly, their work provided an overview of the primary high-throughput experimental screening and characterization methodologies applicable to OSCs, specifically those based on lateral parametric gradients (measuring-intensive) and automated device prototyping (fabrication-intensive). In both scenarios, unequalled rates for the generation of experimental datasets have been achieved that leading to enhancing big data preparedness. Lastly, they used ML algorithms to locate a lucrative application to retrieve quantitative structure-activity connections and extract molecular design reasoning, which is projected to maintain the rate of material discovery in OPV.

Aaron \textit{et al.} \cite{kirkey2020optimization} proposed the design of experiments (DOE) and ML techniques optimizing all-small-molecule OPV cells depending on small-molecule donor, DRCN5T, and non-fullerene acceptors, ITIC, IT-M, and IT-4F. The combination was quick, efficient, and valuable resources enabled sparse but mathematically intentional reasonable sampling of huge parameter spaces. The bulk heterojunction, which is the OPV device's core layer, was optimized in this work. The optimal values of the experimental processing parameters with regard to PCE were then determined using the maps of the PCE landscape that were derived using the ML-based approach for the first and second rounds of optimization. Cagla \textit{et al.} \cite{odabacsi2020machine} discussed the effects of cell manufacturing materials, deposition techniques, and storage conditions on PV cell stability using a dataset containing long-term stability data for 404 organolead halide PSCs. The dataset was created from 181 published papers and analyzed using association rule mining and decision trees-based ML techniques.

Nahdia \textit{et al.} \cite{majeed2020using} proposed an efficient method for analyzing device and material performance incorporating experimental, device modeling, and ML algorithms. The ability to implement manufacturing conditions to device performance by providing a set of electrical device characteristics results in an enlarged and faster improvement of solar energy harvesting devices. Thus, they considered parameters such as annealing temperature, surfactant selection, and charge carrier dynamics in OSCs. Followed by, Bart \textit{et al.} \cite{olsthoorn2019band} presented the predictions related to the bandgap of Organic Crystal Structures with the help of ML techniques. They extracted a consistent dataset of 12,500 crystal structures and the related DFT band gap properties were freely downloaded from a website. The two cutting-edge models combined yield a mean absolute error (MAE) of 0.388 eV, or 13\% of the average band gap of 3.05 eV, for the ensemble. The band gap for 2,60,092 materials in the Crystallography Open Database (COD) is predicted using the trained models.

Fan \textit{et al.} \cite{li2019machine} presented the ML-assisted designing and fabrication of solar cells. The elements can be divided into four sub-categories: Data measurement, material properties, optimization of device architectures, and optimization of fabrication processes. The typical types of ML techniques discussed involve ANN, GA, PSO, SA, RF, etc. Among them, ANN and GA are the two ML techniques that are most frequently used.

%\subsubsection{Output power to be optimised for different light patterns and shading}

%Fabrication techniques implemented for the output power to be optimised for different light patterns and shading.

%---------------------------------------------------------------------------------------------------------------------------------------------------------------------------------

\section{Open research issues and future outlook}
In this section, we highlight some of the key insights that the authors have notably found interesting and consecutively, presents the future outlook of the potential research incorporating ML and the discovery of new materials to develop re-configurable solar cells. In addition, this section also includes the limitations and pitfalls of the ongoing research that needs to be addressed for developing efficient, robust, and stable solar cell architectures. 

%\subsection{Insights of the paper}

%Whilst our search of articles was based on papers published in peer-reviewed journals, accordingly, the following are the key points that we noted in our study. Moreover, this section discusses the key advantages of using ML techniques in the discovery of solar cells. Firstly, our systematic review ML for solar cells targets to review of a total of 58 papers from a total of 18,380. In addition, from the shortlisted papers, we found that the potential of the research work is done in the domain of data-driven approaches for the discovery of solar cells, followed by ML techniques. 

%Moreover, one of the significant results that we obtained from our study classifies the input data, ML algorithms used, and thus, the output expected electrical characteristics of the solar cells. 

According to our review, few articles were published in the domain of using ML for fabricating solar cells. Furthermore, our study revealed that input data was clustered around PSCs, OSCs, and hybrid solar cells. Furthermore, most research used the ANN, GBRT, XGBoost, EXTR, LR, DTR, KNN, RF, SVM, SVR, GPR, and BO algorithms to determine output characteristics such as cost, PCE, the accuracy of the ML model, loss function and error. Lastly, ML was used for optimizing the following solar cell parameters: donor/acceptor ratio, conductivity, donor/acceptor materials, stability optimization, copper content optimization, and spray plasma processing.  

\subsection{Limitations}

Although there are numerous advantages of using ML for solar cell discovery, there are several open issues. From our systematic review, we came across multiple challenges that need to be addressed with regard to the discovery of new  low-cost solar cells. Key among these challenges are:

\begin{itemize}
    \item \textbf{Vulnerability of the input data}. As previously mentioned, most low-cost solar cells were fabricated by trial and error, which leads to high input data vulnerability. \cite{kim2019two} Therefore, model validation should be a necessary step. \cite{trappey2019machine} Moreover, data scarcity is a significant problem in the field of data-driven solar materials science. \cite{sanchez2018inverse} Text mining and picture recognition are considered solutions to overcome these issues of small datasets. \cite{lee2019recent} 
    %Despite the abundance of research articles on data-driven solar materials science and engineering, there are few thorough review papers that are pertinent to this area. \cite{rong2018challenges} The development of new PV technologies will be facilitated by the publication of more comprehensive review articles that outline future directions. \cite{wang2019all} It is noted in the literature that ML algorithms help to expedite the discovery of solar cells. \cite{hachmann2011harvard}

    \item \textbf{Stability of thin-film solar cells}. One of the key concerns in designing low-cost solar cells in the real environment is the stability of organic, inorganic, and hybrid solar cells due to the different compositions of chemical components. These solar cells are very unstable and have a short life period. \cite{ju2018toward} Previously, studies have shown that solar cell efficiency and stability are inversely proportional. Also, the key stability components that need to be addressed are thermal, moisture, and chemical composition stability. \cite{dada2019machine}

    \item \textbf{Inaccurate predictions}. Another key issue with using ML algorithms for discovering solar cells is the inaccurate predictions and outcomes from the ML models. \cite{huuhtanen2018predictive} In most cases, ML algorithms give the confidence interval of the forecasted and predicted values of the solar cells. However, the predicted values for the discovery of solar cells seem to approach up to a maximum of 95\% using the GPR and Bayesian optimization using the probability distribution, which sometimes proves to result in the poor fabrication of solar cells. Therefore, the ML models' prediction models need to be classified properly to avoid such discrepancies. \cite{zhan2016more}

    \item \textbf{Rigorously fabricating solar cells in labs}. The researchers are rigorously fabricating solar cells depending upon the hit and trial methods, which wastes a lot of time, resources, and materials. In addition, if the researchers follow the same procedure in the upcoming years, it is noted that it will further delay the discovery of new materials used to fabricate solar cells. \cite{zhang2019memetic} Moreover, using the permutation and combinations of different layers, electrical characteristics, and other components required to design the solar cells and fabricate solar cells in the laboratory will lead to other consequences which can be avoided with the use of ML techniques and AI integration. \cite{gasparini2019role}

    \item \textbf{Lack of data availability and poor data analysis}. Firstly, it is noted from the study that there is a lack of data availability and, thus, poor data analysis. Second, it is advised to integrate feature engineering, modeling, and domain technical expertise to increase the effectiveness of the created ML model. In parallel, validation experiments should be run to verify the analytical outcomes of the ML model, such as the high-performing prediction candidate. Only a few research studies have used experiments to validate their forecasted materials. \cite{faes2019automated}
\end{itemize}

\subsection{Future outlook}

The future goals and prospective outlook for discovering new low-cost solar cells are mentioned below. Initially, there was a large room for data collection and monitoring to provide input to ML models. Moreover, the extracted data needs feature scaling and data-prepossessing to be used effectively in ML algorithms. Therefore, an appropriate data selection technique must be used to interpolate or extrapolate the data depending on various dependent and independent variables in feature selection. In addition, since ML and AI techniques have recently gained significant importance, adversarial robust ML techniques will play a vital role in forecasting and predicting the design of solar cell architectures. 

Since low-cost solar cell fabrication in a research laboratory is cheap, most researchers tend to retrospectively appreciate the performance of their design after first fabricating the solar cell by trial and error. Instead, we believe it is more beneficial to perform these predictions using robust ML algorithms, which will help design and fabricate more efficient solar cells. Adopting this approach will expedite the solar cell design process. There is also space for research related to the generalized explanations of data extraction and interpretation and to achieve more accurate ML models.  In general, the accuracy of the ML model depends on the input data. Researchers across the globe should target to extract sufficient data and make it available online to help the scientific community discover low-cost, high-performance solar cells. 

\section{conclusions}

In particular, this review covers a broad range of ML techniques for optimizing the performance of low-cost solar cells for miniaturized electronic devices. We reviewed 58 research articles shortlisted from 18,380 research publications found on respective search engines during the past 5 years, which met our inclusion criteria. Moreover, these review articles were shortlisted depending on the four research objectives and, therefore, targeted to answer our defined research questions. According to our systematic review, a majority of research was devoted to data-driven approaches and ML techniques for discovering low-cost solar cells, whereas one-third of all the publications focused on ML algorithms applied to the fabrication process involved. Overall, our systematic review reveals that ML techniques can potentially expedite the discovery of new solar materials and architectures.
Moreover, a new method of classifying the literature has been presented according to data synthesis, ML algorithms, optimization and fabrication process. In addition, the most promising results from our review reveal that the Gaussian Process Regression (GPR) ML technique with the Bayesian Optimization (BO) enables the design of efficient low-cost solar cell architecture. We also presented an outlook into the future goals and prospects for discovering new materials for solar cells.
 
\bibliographystyle{IEEEtran}
\bibliography{ref}

\end{document}